\documentclass[amsmath,twocolumn,aps,pra,reprint]{revtex4-1}
\usepackage{dcolumn}    
\usepackage{bm}         
\usepackage{ifpdf}
\usepackage{amssymb,lineno,amsfonts}
\usepackage{graphicx}   
\usepackage{bbm}
\usepackage{mathrsfs}
\usepackage{upgreek}
\usepackage{epstopdf}
\usepackage{setspace}
\usepackage{hyperref}
\usepackage[matrix,frame,arrow]{xypic}
\usepackage{float}
\usepackage{natbib}
\usepackage{color} 
\newcommand{\ket}[1]{\vert{#1}\rangle}

\newcommand{\outpr}[2]{\vert{#1}\rangle\langle{#2}\vert}

\newcommand{\proj}[1]{\outpr{#1}{#1}}

\newcommand*{\permcomb}[4][0mu]{{{}^{#3}\mkern#1#2_{#4}}}
\newcommand{\comb}[1][-1mu]{\permcomb[#1]{C}}

\begin{document}
	\title{Strong Algorithmic Cooling in Large Star-Topology Quantum Registers}
	\author{$^\dagger$Varad R. Pande, $^\dagger$Gaurav Bhole, $^\dagger$Deepak Khurana, and $^{\dagger,\ddagger}$T. S. Mahesh}
	\email{mahesh.ts@iiserpune.ac.in}
	\affiliation{$^\dagger$Department of Physics and NMR Research center, 
		\\
		$^\ddagger$Center for Energy Sciences,
		\\
		Indian Institute of Science Education and Research, Pune }%

\begin{abstract}
Cooling the qubit into a pure initial state is crucial for realizing fault-tolerant quantum information processing.  
  Here we envisage a star-topology arrangement of reset and computation qubits for this purpose.
    The reset qubits cool or purify the computation qubit by transferring its entropy to a heat-bath with the help of a heat-bath algorithmic cooling procedure.  By combining standard NMR methods with powerful quantum control techniques, we cool central qubits of two large star-topology systems, with 13 and 37 spins respectively.  We obtain polarization enhancements by a factor of over 24, and an associated reduction in the spin temperature from 298 K down to 12 K.  Exploiting the enhanced polarization of computation qubit, we prepare combination-coherences of orders up to 15.  By benchmarking the decay of these coherences we investigate the underlying noise process. Further, we also cool a pair of computation qubits and subsequently prepare them in an effective pure-state.
\end{abstract}
\maketitle


\section{Introduction}
Quantum information processors and other quantum devices are expected to have a major impact on future technologies.  Initializing qubits - the building blocks of a quantum memory,  into a deterministic initial state is an essential process that precedes other quantum operations
\cite{divincenzo2000physical,cirac1999optimal}.  This process is important not only for quantum registers with low-purity initial states, but also for achieving scalable quantum processors with the help of quantum error correction \cite{park2016heat}.
While there are many ways to enhance the purity of a spin-qubit, from optical pumping (eg. \cite{chakraborty2017polarizing}) to exploiting para-hydrogens (eg. \cite{Anwar}), polarization-transfer remains the simplest approach.  Transferring polarization from one spin to another has long been a routine operation in NMR spectroscopy \cite{freeman}.  Inspired by the spin dynamics in these NMR experiments, a universal bound for such entropy transfers in closed systems was calculated by S{\o}rensen \cite{sorensen1989polarization}. 
Later Shulman and Vazirani designed an algorithmic approach for transferring entropy from a smaller set of \textit{computation} qubits, to a larger set of so called \textit{reset} qubits \cite{schulman1999molecular}.  This method, generally known as algorithmic cooling (AC), is essentially a unitary process.
A non-unitary extension of the above pairwise entropy compression between successive qubits was proposed by Boykin \textit{et al.} \cite{boykin2002algorithmic} which resulted in the heat-bath algorithmic cooling (HBAC).  Here, the reset qubits periodically release their excess entropy to a heat-bath so that higher cooling of the computation qubit can be achieved by constructive iterations of AC.
Since then, several HBAC algorithms have been proposed \cite{FERNANDEZ2004,Schulman2005,Elias2006,Schulman2007,Elias2011} and 
numerous experimental studies have also been reported
\cite{chang2001nmr,FERNANDEZ2005,baugh2005experimental,Elias2011126,park2015hyperfine,ryan2008spin,atia2016algorithmic} which use three or five qubit entropy compression circuits as the building block for AC.
More recently asymptotic bounds for HBAC algorithms have been estimated numerically by Raeisi \textit{et al.} \cite{Raeisi} as well as analytically by Rodr\'{\i}guez-Briones \textit{et al.} \cite{Rodr}.
Here we consider HBAC in star-topology quantum registers, which offer highly efficient platforms for this purpose, allowing entropy compression between the computation qubit and a large number of reset qubits.  

In the following section, we first describe star-topology systems and then explain HBAC procedure in such systems.  Subsequently, in section III, we describe NMR experiments of HBAC on two star-topology quantum registers consisting of 13 and 37 spins respectively.  As an application of HBAC, we report the preparation of large combination coherences in section III B. We have also reported benchmarking the decay of these coherences and attempt to understand the dominant source of noise.
Before we summarize, we also explain, in section III C, the sequential HBAC of a pair of computation qubits, and subsequently preparing them into an effective pure state.

\section{Theory}
\subsection{Star-topology quantum register}
\begin{figure}
	\centering
	\includegraphics[trim=0.6cm 0.4cm 0.9cm 0cm, clip=true,width=7.5cm]{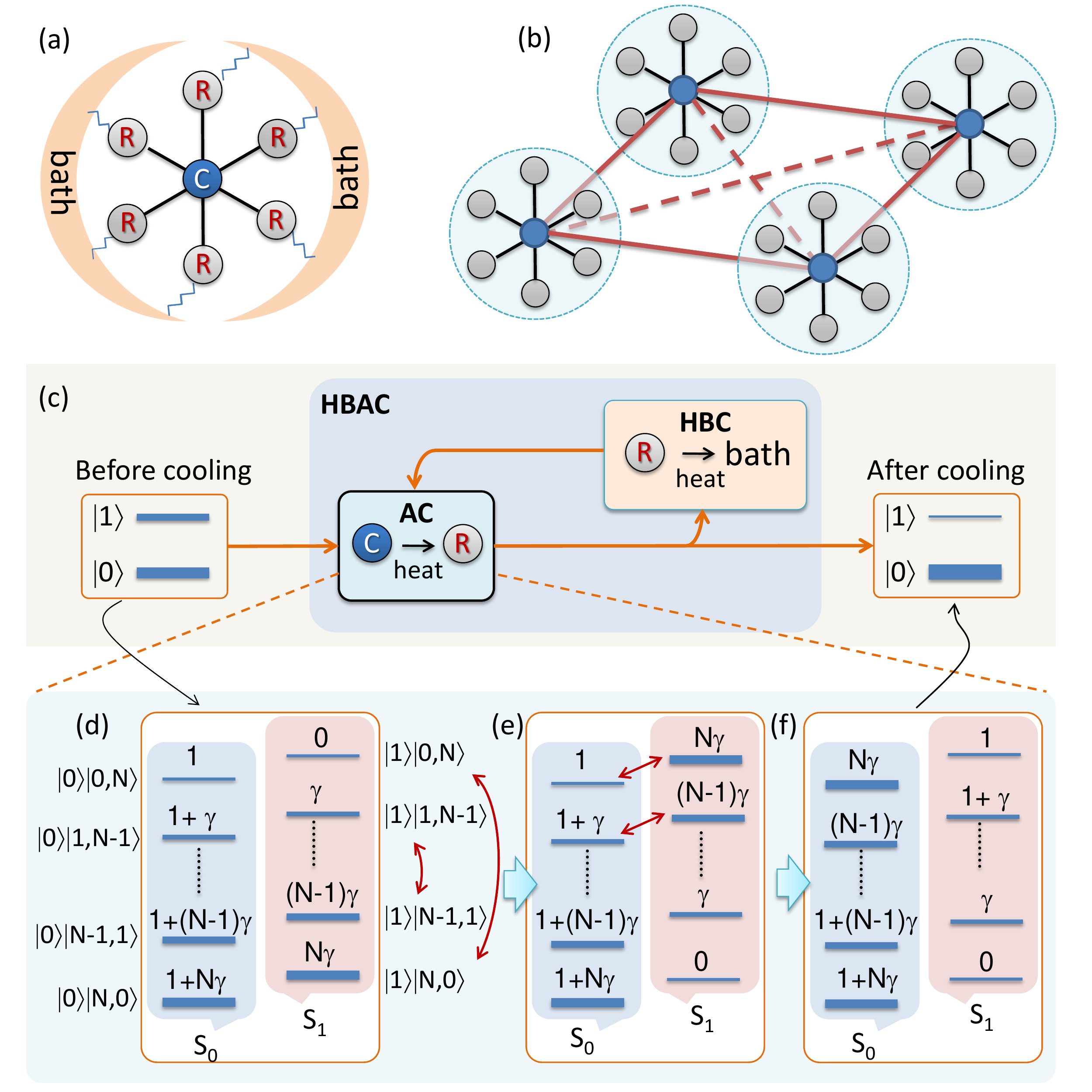}
	\caption{(a) Star-topology with a central computation qubit and a set of reset qubits surrounded by a bath, (b) a four-star quantum register, (c) HBAC procedure.  The different steps of AC are illustrated in (d-f).  In (d) the various levels are labeled as $\ket{\alpha}\ket{N-j,j}$, where $\alpha = \{0,1\}$ corresponds to the two states of the computation qubit and $j \in \{0,\dots,N\}$  denotes the number of reset qubits in state $\ket{1}$.}
	\label{star}
\end{figure}

In star-topology, a single \textit{computation} qubit ({\tt C}) is uniformly coupled to a set of $N$ identical (magnetically equivalent) \textit{reset}  qubits ({\tt R}) with the same interaction strength $J_{\tt RC}$ (Fig. \ref{star}).  The interactions among the reset qubits are assumed to be either negligible or, as in our case, ineffective due to magnetic equivalence symmetry. Star-topology quantum systems have already found several interesting applications such as in field sensing \cite{Jones1166}, spectroscopic measurements \cite{shukla2014noon}, as well as for understanding noise in quantum systems \cite{khurana2016spectral}.

In the NMR setting described here, the qubits are formed by spin-1/2 nuclei of a molecular ensemble placed in a strong static magnetic field $B_0 \hat{z}$.  For effectiveness, we choose the computation and reset qubits to be of different nuclear isotopes with  Larmor frequencies $\omega_0^{\tt C} = -\gamma_{\tt C} B_0$ and 
$\omega_0^{\tt R} = -\gamma_{\tt R} B_0$ respectively, where $\gamma_{\tt C}, \gamma_{\tt R}$ are the gyromagnetic ratios such that $\gamma = \gamma_{\tt R}/\gamma_{\tt C} \ge 1$.  The Hamiltonian of such a star-system in a doubly rotating frame is 
\begin{eqnarray}
	{\cal H} = -\frac{\hbar\omega_{\tt C}}{2} \sigma_z^{\tt C} -  \frac{\hbar\omega_{\tt R}}{2} \sum_{j=1}^N \sigma_{jz}^{\tt R} +
	\frac{\hbar\pi J_{\tt RC}}{2} \sum_{j=1}^{N} \sigma_z^{\tt C}\sigma_{jz}^{\tt R},
\end{eqnarray}
where $\sigma_z$ represents the Pauli operator and $\omega_{\tt C}$, $\omega_{\tt R}$ are tunable resonance off-sets which can be set to zero without loss of generality.

Once cooled, the computation qubit should ideally remain in the low-entropy state for a long time.  To that end, it needs to be sufficiently isolated from the bath and thereby have a long memory in terms of its spin lattice relaxation time $\mbox{T}_1^{\tt C}$. On the other hand, the reset qubits need to strongly interact with the bath, and should ideally have a short $\mbox{T}_1^{\tt R}$ so that many cycles of HBAC can be performed.  This requirement is very well satisfied by the star-topology systems  wherein the outer reset qubits seem to shield the central computation qubit from the environmental influences.

At high-temperature   ($kT \gg \hbar \omega_{\tt C}$) limits, the thermal-equilibrium density matrix $\rho_\mathrm{eq}^{\tt C}$ of  the computation qubit can be approximated as a convex sum of the maximally mixed state ($\mathbbm{1}/2$) and a \textit{deviation} ($\rho_\Delta^{\tt C}$), i.e., 
\begin{eqnarray}
	\rho_\mathrm{eq}^{\tt C}    
	= \frac{1}{2} \exp\left(\frac{\hbar \omega_{\tt C} \sigma_z^{\tt C}}{2kT}\right)
	&\approx&  (1-\epsilon) \frac{\mathbbm{1}}{2} + \epsilon \proj{0_{\tt C}}
\end{eqnarray}
in the eigenbasis $\{\ket{0_{\tt C}},\ket{1_{\tt C}}\}$ of the Zeeman Hamiltonian.
Here the dimensionless quantity $\epsilon =  \hbar \omega_{\tt C}/(2kT)$ is a measure of the purity of the thermal state.  

Spin temperature $T^{\tt C}$ for an arbitrary single-qubit state $\rho^{\tt C}$, with diagonal elements $\rho_{00}^{\tt C}$ and $\rho_{11}^{\tt C}$ and null off-diagonal elements in the Zeeman basis, is quantified by the Boltzmann distribution, 
\begin{eqnarray}
	\frac{\rho_{11}^{\tt C}}{\rho_{00}^{\tt C}} = \exp \left[ -\frac{\hbar \omega_0^{\tt C}}{kT^{\tt C}}
	\right].
	\label{st_c}
\end{eqnarray}
While $T^{\tt C} = T$ at thermal equilibrium, the objective is to \textit{cool} the computation qubit such that $T^{\tt C} \ll T$ by redistributing $\rho_{00}^{\tt C}$ and $\rho_{11}^{\tt C}$.
As shown in Fig. \ref{star} (c), we can achieve cooling by transferring the magnetization iteratively from reset qubits to the computation qubit.  Each iteration involves two stages - (i) algorithmic cooling (AC), wherein entropy is transferred from the computation qubit to the reset qubit, and (ii) heat-bath cooling (HBC), wherein the reset qubits give away the extra entropy gained to the bath.  Together these stages form heat-bath algorithmic cooling (HBAC).  

\subsection{HBAC in Star-topology Quantum Registers}
First we explain AC in star-registers which involves a two-step procedure.  The $2^{N+1}$ energy-levels of the $N+1$ star-register,  can be grouped into two subspaces $S_0$ and $S_1$ corresponding to $\ket{0}$ and $\ket{1}$ states of the computation qubit (see Fig. \ref{star} (d)), such that the deviation matrix
\begin{eqnarray}
\rho_\Delta = \proj{0_{\tt C}}  \otimes S_0 + \proj{1_{\tt C}} \otimes S_1.
\end{eqnarray}
In each subspace  $S_\alpha$, the star-symmetry imposes degeneracy in all the levels except in the ground and the most-excited levels.
Degeneracy of the level $\ket{\alpha}\ket{N-j,j}$ in Fig. \ref{star} (d) is $^NC_j = N!/\{j!(N-j)!\}$.

The two subspaces $S_0$ and $S_1$ are identical, except that $S_1$
has a higher energy and therefore a lower  population.  
The first step in AC involves the population inversion of $S_1$ (see Fig. \ref{star} (d) and (e)). It can be realized experimentally by applying a transition-selective $\pi$ pulse on the common $+J/2$ transition of the reset qubits.  The second-step involves inverting top $m$ transitions of the computation qubit 
(see Fig. \ref{star} (e) and (f)).  
One might naively choose $m = \lfloor N/2 \rfloor$, the largest integer less than or equal to $N/2$.  However, numerical calculations suggest that there exists an optimum $m$ for a given star-system (Appendix \ref{appA}).  Of course, it also depends on the total number as well as the fidelity of HBAC iterations. 
These two operations of AC drive a large population from $S_1$ to $S_0$ and in effect transfer the entropy to reset qubits.  The next step is HBC which simply involves a delay $\tau_\mathrm{HB}$ that is just long enough for the reset qubits to release the excess entropy to the heat bath.  An excess delay will undesirably allow the computation-qubit to heat-up again (Appendix \ref{appA4}).  AC and HBC together form one iteration of HBAC.  After sufficiently many HBAC iterations, the reset qubits are traced out, and the cooled computation qubit can be used for further processing.

\subsection{Estimation of Cooling}
We now describe numerical estimation of HBAC efficiency in a star-system.  The dynamics of relative populations and hence that of the longitudinal magnetization of a pair of levels with a  spin-lattice relaxation time $\mbox{T}_1$ follows Bloch's equation \cite{bloch1946nuclear}.  However, for a star-system with $(N+1)$-qubits, tracking all the populations ($2^{N+1}$)  appears to be a daunting task.  Nevertheless, by exploiting the symmetry of the star-system as well as the time-scale separation between the memories of ${\tt C}$ and ${\tt R}$ qubits  (i.e., $\mbox{T}_1^{\tt C} \gg \mbox{T}_1^{\tt R}$), we can analyze the population dynamics and calculate the relative magnetization $M_n$ and the corresponding spin temperature $T^{\tt C}_n$ after $n^{\mathrm{th}}$ iteration (Appendix \ref{appA}).
In NMR, $kT^{\tt C}_n \gg \hbar \omega^{\tt C}_0$, and therefore 
\begin{eqnarray}
T^{\tt C}(n) \simeq T/M_n,
\end{eqnarray}
where we have set $M_0 = 1$ as the thermal equilibrium magnetization at the sample temperature $T$. 

\begin{figure}
	\centering
	\includegraphics[trim=2.5cm 4.6cm 2.5cm 0cm, clip=true,angle=0,width=8cm]{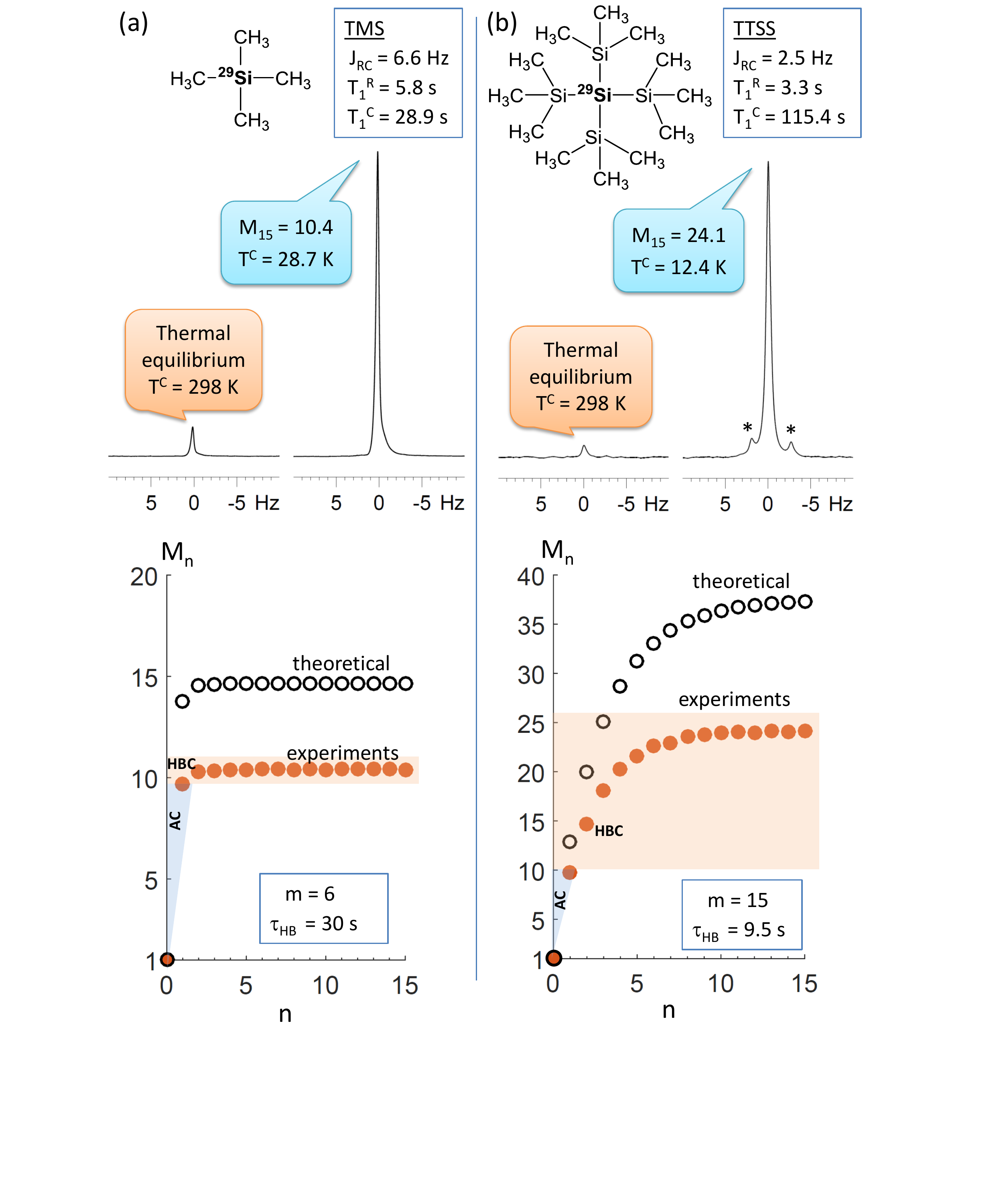}
	\caption{HBAC of (a) TMS and (b) TTSS.  Shown in each case are - molecular structure, $^1$H-decoupled $^{29}$Si spectra before and after HBAC, and magnetization versus HBAC iteration number ($n$).  Here the theoretical data-points are calculated with ideal HBAC controls.   In the right spectrum, the sidebands (indicated by stars) are due to $^{29}$Si--$^{13}$C J-coupling. }
	\label{expt1}
\end{figure}

\section{Experiments}
\subsection{HBAC in 13-star and 37-star systems}
We performed HBAC in
(i) 13-star system: tetramethylsilane (TMS) (Fig. \ref{expt1} (a)) as well as (ii) 37-star system: tetrakis(trimethylsilyl)silane (TTSS) (Fig. \ref{expt1} (b)), each dissolved in CDCl$_3$.
In both cases, all the $^1$H spins act as reset qubits and the central naturally-abundant $^{29}$Si spin acts as the computation qubit.
The coupling constants ($J_{\tt RC}$), T$_1$ values, the number of swapped energy levels ($m$), and heat-bath durations ($\tau_{\mathrm{HB}}$) used in the experiments are displayed in Fig. \ref{expt1}. 
All the experiments were carried out in a 400 MHz Bruker NMR spectrometer at an ambient temperature of 298 K.

The subspace inversion in the first part of AC (see Fig. \ref{star} (d)) was achieved by a transition-selective Gaussian-shaped $\pi$ pulse of duration 750 ms which was on-resonant on the $^1$H transition corresponding to $S_1$ subspace.  The second part of AC involving inter-subspace swapping of top $m$ energy-levels (see Fig. \ref{star} (e)) was achieved by an amplitude and phase modulated RF pulse which inverts only the transitions with positive frequencies [Appendix \ref{appB}].
The heat-bath delay $\tau_\mathrm{HB}$ was set to 30 s and 9.5 s for TMS and TTSS respectively. Finally, we measured $^{29}$Si magnetization ($M_n$) from the $^1$H decoupled $^{29}$Si spectra recorded versus the HBAC iteration-number ($n$).  
The experimental results are compared with numerical predictions in Fig. \ref{expt1}.  Representative $^{29}$Si spectra comparing before and after HBAC are also shown  in Fig. \ref{expt1}.  In the case of TMS, the magnetization after 15 HBAC iterations was enhanced by a factor of 10.4, which corresponds to a spin temperature of 28.7 K.  In the case of TTSS, the enhancement factor after 15 HBAC iterations was 24.1, which   corresponds to a spin temperature of 12.4 K.

The numerically predicted upper-bounds for the magnetizations under perfect HBAC iterations are also shown in Fig. \ref{expt1}.
Comparison of the theoretical limits with the experimental results suggests a scope for further enhancement of cooling.  The lower values of experimentally obtained cooling are mainly due to finite fidelity of GRAPE pulses, RF inhomogeneities, decoherence, and nonlinearities in the spectrometer hardware. 
Quantitative details on the imperfections are described in appendix (Appendix \ref{appA2}). 

\begin{figure}
	\centering
	\includegraphics[trim=0cm 0cm 7.5cm 0.1cm, clip=true,angle=0,width=7.5cm]{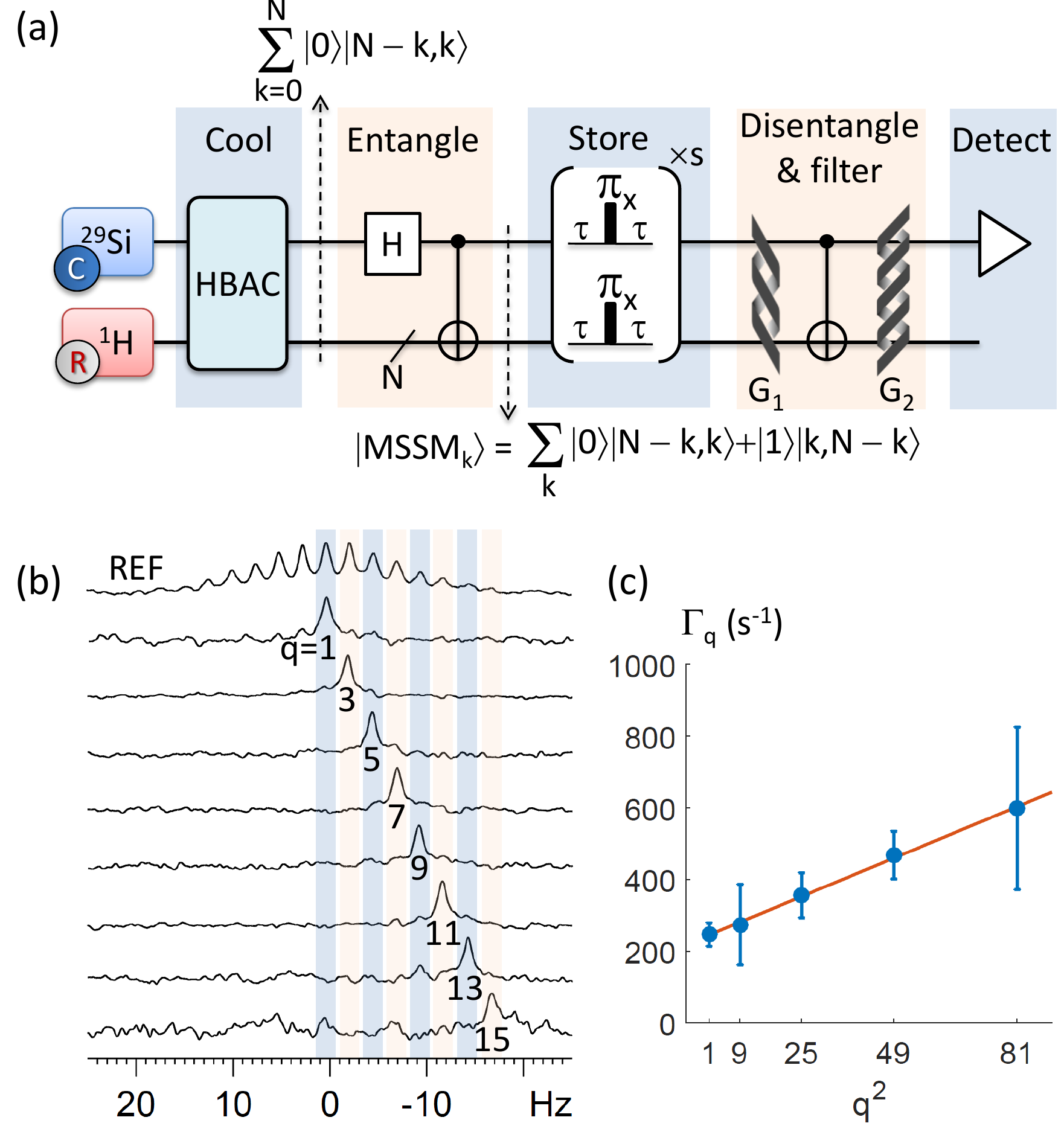}
	\caption{(a) Quantum circuit for preparing combination coherences.  Here due to the star-symmetry, each CNOT gate acts simultaneously on all the reset qubits.  A pair of pulsed-field-gradients G$_1$ and G$_2$  select a coherence of order-$q$ which is  converted into an observable single-quantum coherence by the second CNOT \cite{shukla2014noon}.
		(b)	$^{29}$Si reference spectrum of TTSS and characteristic spectra corresponding to various coherences of odd orders $q$. (c) The decay rates ($\Gamma_q$) versus $q^2$ measured in TTSS.}
	\label{mqss}
\end{figure}

\subsection{Achieving large combination-coherences}
As an immediate application of the strong polarization obtained by HBAC, we prepare and filter-out large combination coherences \cite{levitt2001spin}, which are essential in realizing a scalable quantum processor. 
The quantum circuit for studying such coherences is shown in Fig. \ref{mqss} (a).
For a star-system with  even-$N$, the $q$-quantum combination coherence appears as a single transition with a frequency $(q-1)J_{\tt RC}/2$ from the central transition.  The experimental spectra corresponding to odd-quantum coherences with orders up to $15$ are displayed in  Fig. \ref{mqss} (b). 

Further, as indicated by the circuit in Fig. \ref{mqss} (a), we benchmark the life-time of various coherences (with orders from $q=1$ to $9$) under the  CP storage \cite{carr1954effects} with an inter-pulse delay $2\tau = 500~ \upmu$s.  Tang and Pines \cite{tang1980multiple} had earlier predicted that in the case of completely correlated noise, the adiabatic term of the relaxation rate $\Gamma_q$ originated from the energy-conserving processes is proportional to the square of the coherence order, i.e.,
$\Gamma_q \propto q^2$.  Fig. \ref{mqss} (c) displays a good linear fit of $\Gamma_q$ versus $q^2$ for TTSS.  It is  interesting to note that the noise in TTSS is hence predominantly correlated.  It may also be noted that benchmarking in a 12-qubit system was earlier reported by Negrevergne et al \cite{negrevergne2006benchmarking}.

\begin{figure}
	\centering
	\hspace*{-0.5cm}
	\includegraphics[trim=4cm 1.4cm 6cm 0.5cm, clip=true,angle=0,width=7cm]{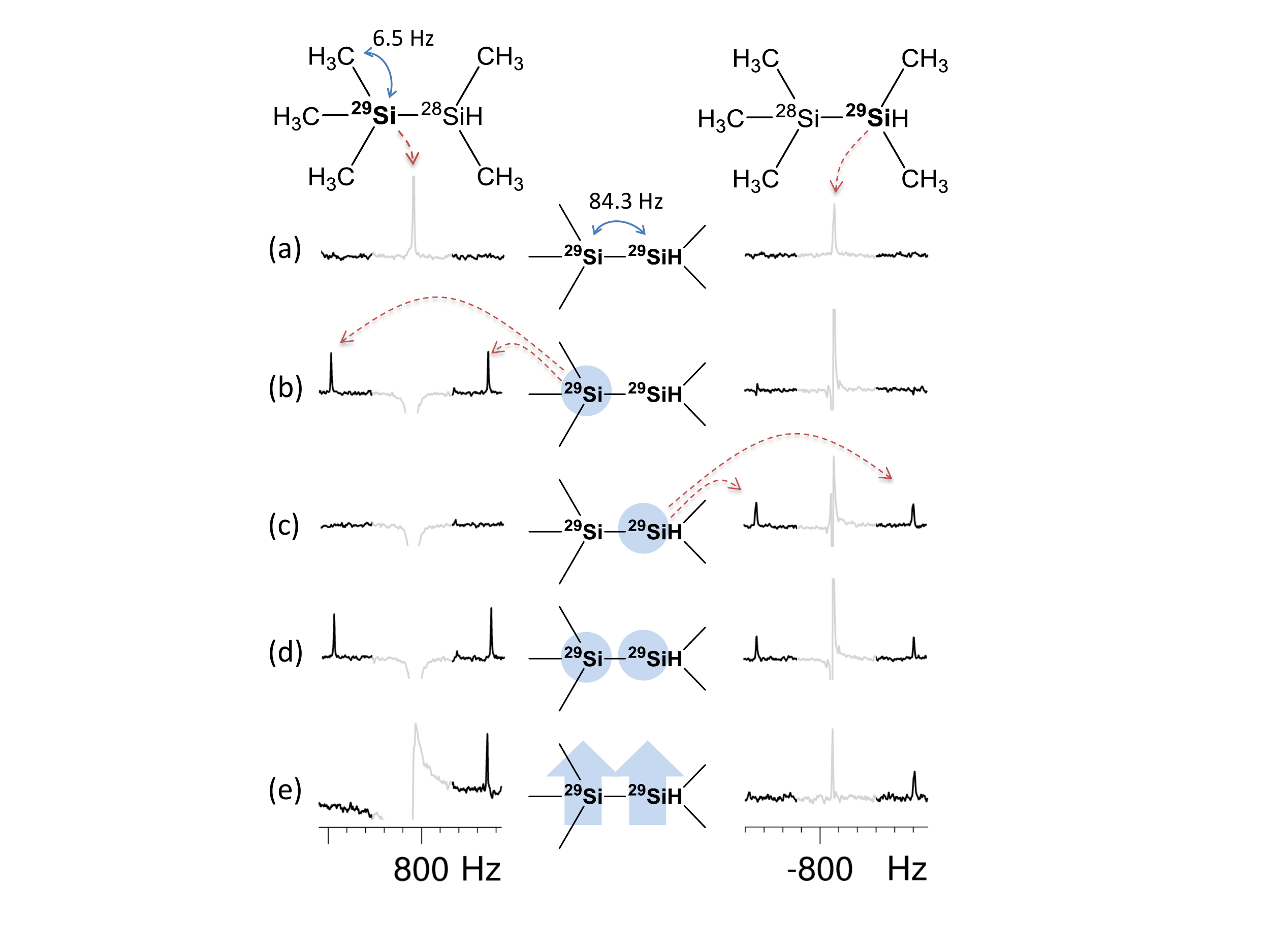}
	\caption{HBAC of a pair of computation qubits.  (a) The reference spectra of $^{29}$Si corresponding to the thermal equilibrium magnetization. Spectra obtained after (b) cooling the left qubit, (c) transferring the polarization from left qubit to right qubit, (d) cooling the left qubit again, and (e) finally preparing the two $^{29}$Si qubits in the $\ket{00}$ pseudopure state.}
	\label{twoq}
\end{figure}

\subsection{A pair of computation qubits}
We now describe HBAC of a pair of computation qubits using two naturally abundant $^{29}$Si spins of pentamethyldisilane (PDS) dissolved in CDCl$_3$ (see Fig. \ref{twoq}).  The RF off-set was chosen such that the resonance frequencies of the two $^{29}$Si spins were $\pm 806$ Hz,
and their $T_1$ values are 47.2 s and 39.5 s respectively. $T_1$ values of $^1$H spins are about 8.5 s.
The coupling between the methyl protons and the closest $^{29}$Si was 6.5 Hz while that between the two $^{29}$Si spins was 84.3 Hz.   

The spectra corresponding to thermal equilibrium magnetization are shown in Fig. \ref{twoq} (a).  The two central peaks (de-emphasized) at $\pm 806$ Hz correspond to those molecules wherein only one of the silicons is $^{29}$Si, while the other being NMR-inactive $^{28}$Si (spin-0 nucleus).  Here, no  signal from a pair of interacting qubits ($^{29}$Si-$^{29}$Si) can be observed.  Fig. \ref{twoq} (b) displays the spectra obtained after algorithmic cooling (AC) of the left qubit, while the right qubit remains unobservable.  The pair of spectral lines in the left qubit is due to splitting caused by $^{29}$Si-$^{29}$Si J-coupling.  We then transferred the polarization from left qubit to right qubit using a SWAP gate. As evident in the spectra displayed in Fig. \ref{twoq} (c), the pair of lines in the left  disappear, while the corresponding pair in the right appear.  After a heat-bath delay of 20 s, we carryout a second algorithmic cooling of the left qubit, so that both the qubits are now polarized as indicated by Fig. \ref{twoq} (d).  Finally we prepare $\ket{00}$ pseudopure state of the two cooled qubits using spatial averaging technique \cite{cory1997ensemble}.  The pair of qubits are now ready for implementing quantum gates or any other algorithms.  

\section{Summary and discussions}
We proposed an arrangement of star-topology quantum registers that allows cryorefrigeration of spin-qubits by drastically reducing the spin temperature. Here each of the computation qubits is surrounded by a symmetric set of reset-qubits.  We described the heat-bath algorithmic cooling of computation qubits by iterating a two-step procedure of transferring entropy from the computation qubits to the reset qubits and subsequently to a heat bath.  Using NMR techniques, we experimentally demonstrated the heat-bath algorithmic cooling in two different star-systems of 13 and 37 spins respectively.  We obtained strong polarization enhancements up to a factor of over 24 which reduces the spin-temperature from room-temperature down to 12 K.  Using a $^{29}$Si qubit with a natural abundance of mere 4.6\% as the computation qubit, we were able to prepare combination quantum coherences of orders up to 15.  By  measuring the life-times of these coherences we studied the noise characteristics in the 37-spin system.  Further, on another system having a naturally abundant pair of $^{29}$Si, we demonstrated a sequential heat-bath algorithmic cooling and subsequently prepared a pseudopure state.  

Star-topology registers promise interesting applications in a variety of quantum devices wherein robust cooling of computation qubits is necessary.  In an addressable array of star-registers, it is possible to achieve further cooling by  incorporating the above method with other entropy compression schemes. The present work may also have ramifications in the spectroscopy of low natural-abundance nuclear isotopes such as $^{29}$Si demonstrated in this work. 
In principle, it is possible to combine the method described here with hyperpolarization techniques to achieve higher cooling.  Further, by using a stronger magnetic field and a sophisticated detection hardware such as a cryo-probe one can prepare and detect higher coherence orders.

\begin{acknowledgments}
Authors acknowledge V. S. Anjusha and Sudheer Kumar for discussions.
This work was partly supported by DST/SJF/PSA-03/2012-13 and CSIR 03(1345)/16/EMR-II. VRP and GB acknowledge support from DST-INSPIRE fellowship.
\end{acknowledgments}

\appendix

\section{Numerical estimation of HBAC in star-topology registers
\label{appA}	
	}

\subsection{Population SWAPs}
\label{appA1}
The traditional way to treat the relaxation in a multi-level system is via Solomon equations \cite{solomon1955relaxation}.  However,
here we provide a simpler analysis by exploiting the symmetry of the star-system as well as the fact that  reset qubits have much shorter memory than the computation qubit, i.e., $\mathrm{T}_1^{\tt R} \ll \mathrm{T}_1^{\tt C}$.  This condition is very well satisfied by the experimental spin-systems described in the text.  The time-scale separation implies the following: on disturbing the thermal equilibrium state, the intra-subspace transitions quickly re-establish Boltzmann distribution within each subspace, while the inter-subspace transitions take place at a much slower rate.
Although the thermal equilibrium is established eventually, we are interested in the intermediate time scales at which the inter-subspace population difference is maximized, and hence higher magnetization is achieved.

Consider a subspace $S_\alpha$ of Fig. \ref{star} (d) of the main text.  
Consistent with the high-temperature Boltzmann distribution,
the equilibrium populations of $j$th eigenstates in $S_0$ and $S_1$ are
set to (up to a scaling factor)
\begin{eqnarray}
p_j^0 = b_j ~~ \mbox{and} ~~ p_j^1 = b_j-1
\end{eqnarray}
respectively, where $b_j = b+1+\gamma(N-j)$ with 
$b$ being the uniform background population. 
Thus, up to an uniform background and a scaling, the population difference between a pair of levels connected by reset-qubit transition (intra-subspace) is $\gamma$, while that connected by a computation qubit transition (inter-subspace) is normalized to be 1.  
The populations of $\comb{N}{j}$-fold degenerate levels $\ket{0}\ket{N-j,j}$ and $\ket{1}\ket{N-j,j}$ are
\begin{eqnarray}
B_j^0 &=& \comb{N}{j}  b_j
~\mbox{and}  \nonumber \\
B_j^1 &=&    \comb{N}{j} (b_j-1),
\label{pj}
\end{eqnarray}
and $B^0 = \sum_j B_j^0$, $B^1 = \sum_j B_j^1$ are the total populations of the two subspaces respectively.
The thermal equilibrium magnetization $M_0$ of the computation qubit is proportional to the inter-subspace population difference, i.e., 
\begin{eqnarray}
M_0 = m_{\tt C} (B^0-B^1) =  m_{\tt C}\sum_j \comb{N}{j} = m_{\tt C} 2^N,
\end{eqnarray}
where $m_{\tt C}$ is a proportionality constant.  Without loss of generality, we can set $M_0=1$.

The first step of the algorithmic cooling involves the  population-inversion of $S_1$, while the second step involves swapping the top $m < \lfloor N/2 \rfloor$ transitions (here $\lfloor x \rfloor$ indicates the greatest integer less than or equal to $x$) of the computation qubit.
Effectively, it results in the population swapping of top-half of $S_0$ with bottom-half of $S_1$,
i.e.,
\begin{eqnarray}
p_j^0 &\stackrel{\eta_j}{\longrightarrow}&
(1-\eta_j)p^0_j + \eta_j p^1_{N-j}, ~~\mbox{and}~~\nonumber \\
p_{N-j}^1 &\stackrel{\eta_j}{\longrightarrow}&
(1-\eta_j)p^1_{N-j} + \eta_j p^0_j,
\label{swap}
\end{eqnarray}
where $\eta_j$ denotes swapping factor. 
Although ideally
\begin{eqnarray}
\begin{array}{c|l}
\eta_j =& 1 ~\mbox{for}~ j >  N-m \nonumber \\
& 0 ~\mbox{otherwise},
\end{array}
\end{eqnarray}
imperfections of practical SWAP gates  lead to lower degree of algorithmic cooling. 
The populations of various levels of either subspaces (at time t=0 indicated by the parenthesis) are now
\begin{eqnarray}
p_j^0(0) &=& (1-\eta_j)b_j+\eta_j (b_{N-j}-1), ~ \mbox{and} \nonumber \\
p_{N-j}^1(0) &=& (1-\eta_j)(b_{N-j}-1)+\eta_j b_j.
\end{eqnarray}

Denoting the initial subspace populations by
\begin{eqnarray}
P^0(0) &=& \sum_j \comb{N}{j} ~ p_j^0(0), ~ \mbox{and} \nonumber \\
P^1(0) &=& \sum_j \comb{N}{j} ~ p_j^1(0),
\end{eqnarray}
the relative magnetization of computation qubit for the first cooling iteration
\begin{eqnarray}
M_1 = \frac{P^0(0)-P^1(0)}{B^0-B^1}.
\end{eqnarray}

If now a heat bath delay is introduced, the inter-subsystem transitions slowly drive the computation qubit towards thermal equilibrium via Bloch equation, i.e.,
\begin{eqnarray}
M(\tau) = 1+(M_1-1 )
e^{-\tau/\mbox{T}_1^{\tt C}}.
\label{mtau}
\end{eqnarray}
For the second iteration, before we can calculate the effect of SWAP operation (Eq. \ref{swap}), we need to know the populations of various levels in each subspace.  This is described in the following.

\begin{figure}[b]
	\centering
	\hspace*{-0.2cm}
	\includegraphics[trim=2cm 2cm 2cm 3cm, clip=true,angle=0,width=7cm]{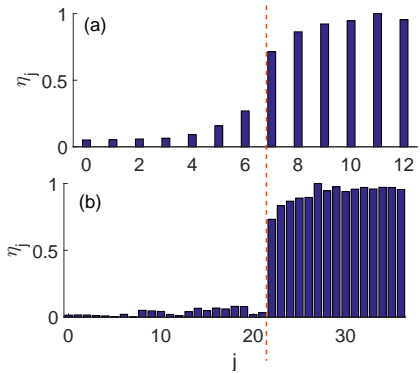}
	\caption{Estimated swapping factors $\eta_j$ versus the transition number $j$ of the computation qubit in (a) TMS and (b) TTSS as indicated.}
	\label{eta}
\end{figure}

\subsection{Intra-subspace relaxation}
\label{appA2}
For the moment, we ignore the inter-subspace relaxation which occurs at a much slower rate.  For simplicity, let us consider only $S_0$ subspace, but similar calculations hold for $S_1$ as well.
Consider a pair of energy levels $(j,j+1)$ connected by the reset qubit transition with energy $\hbar \omega_R$ and prepared with initial populations $p_{j+1}^0(0)$ and $p_j^0(0)$.
The initial magnetization is proportional to
the population difference $\Delta p^0_{j,j+1}(0) = \{p_j^0(0) - p_{j+1}^0(0)\}$, i.e.,
\begin{eqnarray}
M_{\tt R}^{j,j+1}(0) = m_{\tt R} \Delta p_{j,j+1},
\end{eqnarray}
and it gradually evolves towards the equilibrium magnetization $m_{\tt R}\gamma $ via Bloch equation 
such that the instantaneous magnetization
\begin{eqnarray}
M_{\tt R}^{j,j+1}(\tau) = m_{\tt R} \gamma + [M_{\tt R}^{j,j+1}(0)-m_{\tt R} \gamma]e^{-\tau/\mbox{T}_1^{\tt R}}. ~~~
\end{eqnarray}
Thus, in terms of populations we obtain,
\begin{eqnarray}
p^0_{j}(\tau)-p^0_{j+1}(\tau) = \gamma + 
\{\Delta p^0_{i,i+1}(0)-\gamma \}
e^{-\tau/\mbox{T}_1^{\tt R}},~~~
\end{eqnarray}
where all terms in the right hand side are already known.  Thus, in the multi-level system with $(N+1)$-unknowns $p_j(\tau)$, we 
obtain a set of $N$ coupled linear equations.  Solving these equations together with the overall population conserving equation 
\begin{eqnarray}
\sum_j \comb{N}{j} ~ p^0_j(\tau) = \sum_j \comb{N}{j} ~ p^0_j(0),
\end{eqnarray}
we can determine all the temporal populations in $S_0$.  Similarly, we can determine the populations in $S_1$.

\subsection{Inter-subspace relaxation}
\label{appA3}
We assume that all the transitions of the computation qubit have the same relaxation time constant $\mbox{T}_1^{\tt C}$. In that case, using Eq. \ref{mtau}
\begin{eqnarray}
p_j^0(\tau)-p_j^1(\tau) = 1 + \left\{
p_j^0(0)-p_j^1(0)-1
\right\} e^{-\tau/\mbox{T}_1^{\tt C}}.
\end{eqnarray}
Using above equation along with 
$p_j^0(\tau)+p_j^1(\tau) = p_j^0(0)+p_j^1(0)$,
we can determine $p_j^0(\tau)$ and $p_j^1(\tau)$. 

We can now calculate the effect of SWAP (Eq. \ref{swap}) and thereby estimate the relative magnetization $M_2$ for the 2nd iteration of HBAC.  In addition, by interpolating the experimentally obtained magnetizations $M_n$ we can also estimate the swapping factors $\eta_j$.  Fig. \ref{eta} shows the swapping factors for both TMS and TTSS.  

\subsection{Variation of $M_{15}$ with $N$, $m$ and $ \tau_{HB} $}
\label{appA4}
\begin{figure}[h]
	\hspace*{-0.2cm}
	\includegraphics[trim=1cm 1cm 1cm 1cm, clip=true,angle=0,width=6.5cm]{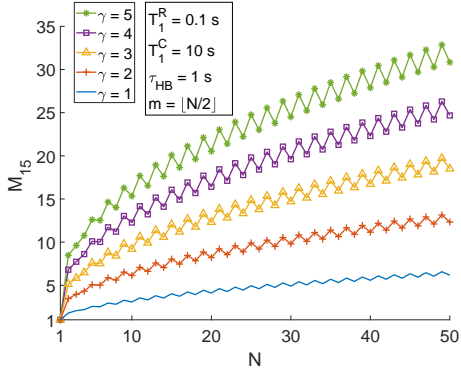}
	\caption{Simulated magnetizations after 15 iterations of HBAC versus star-size ($N$) for a set of relative gyromagnetic ratios $(\gamma)$.  The HBAC parameters are shown in the inset.}
	\label{predictM}
\end{figure}

In the following we describe some interesting scenarios as predicted by the above model of HBAC in star-topology systems.  First of all, for a given set of HBAC parameters ($\mbox{T}_1^{\tt R}$, $\mbox{T}_1^{\tt R}$, $\tau_{HB}$) we find a zig-zag pattern for magnetization ($M_n$) as a function of size of the star ($N$).  Fig. \ref{predictM} displays $M_n$ versus $N$ curves for a range of $\gamma$ values.  Interestingly, HBAC efficiency is higher for odd values of $N$.  Of course, these profiles depend on the number of high-energy transitions (of computation qubit) which are being swapped.  For a given system, it might be possible to optimize this number for various iterations to achieve the best cooling rate.

\begin{figure}[h]
	\hspace*{-0.2cm}
	\includegraphics[trim=1.5cm 4cm 0.5cm 4cm, clip=true,angle=0,width=8.5cm]{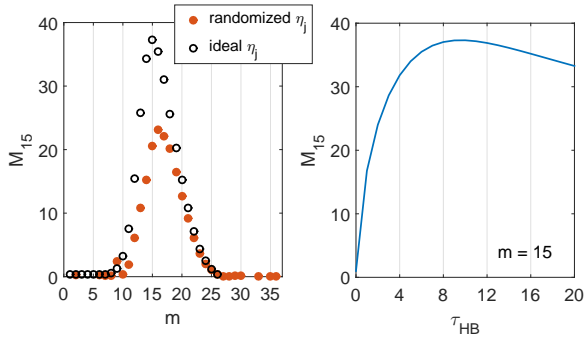}
	\caption{(a) Simulated  magnetization of TTSS versus the number of top-levels swapped ($m$) during HBAC iterations.  In the case of randomized $\eta_j$ values, $\eta_j = 0$ and $\eta_j = 1$ are replaced with distributions $[0,0.2]$ and $[0.8,1]$ respectively. 
		(b) Magnetization versus the heat-bath delay.  
	}
	\label{predictTauNTR}
\end{figure}

Secondly, we also show the dependence of magnetization on the heat-bath delay $\tau_{HB}$.  Fig. \ref{predictTauNTR} (b) illustrates this dependence for a particular set of parameters.  As expected, there is an optimum delay at which maximum cooling can be achieved.

Another interesting optimization parameter is $m$, the number of transitions of the computation-qubit to be inverted during HBAC iterations.  Fig. \ref{predictTauNTR} (a) displays the numerical simulation of $M_{15}$ versus $m$.  It appears that for a system with finite memories ($\mbox{T}_1^{\tt R}$, $\mbox{T}_1^{\tt C}$),  generally $m < \lfloor N/2 \rfloor$.  In fact, one can also vary the $m$ value during the HBAC iterations.

\begin{figure}
	\hspace*{-0.2cm}
	\includegraphics[trim=4cm 0cm 5cm 0cm, clip=true,angle=0,width=8cm]{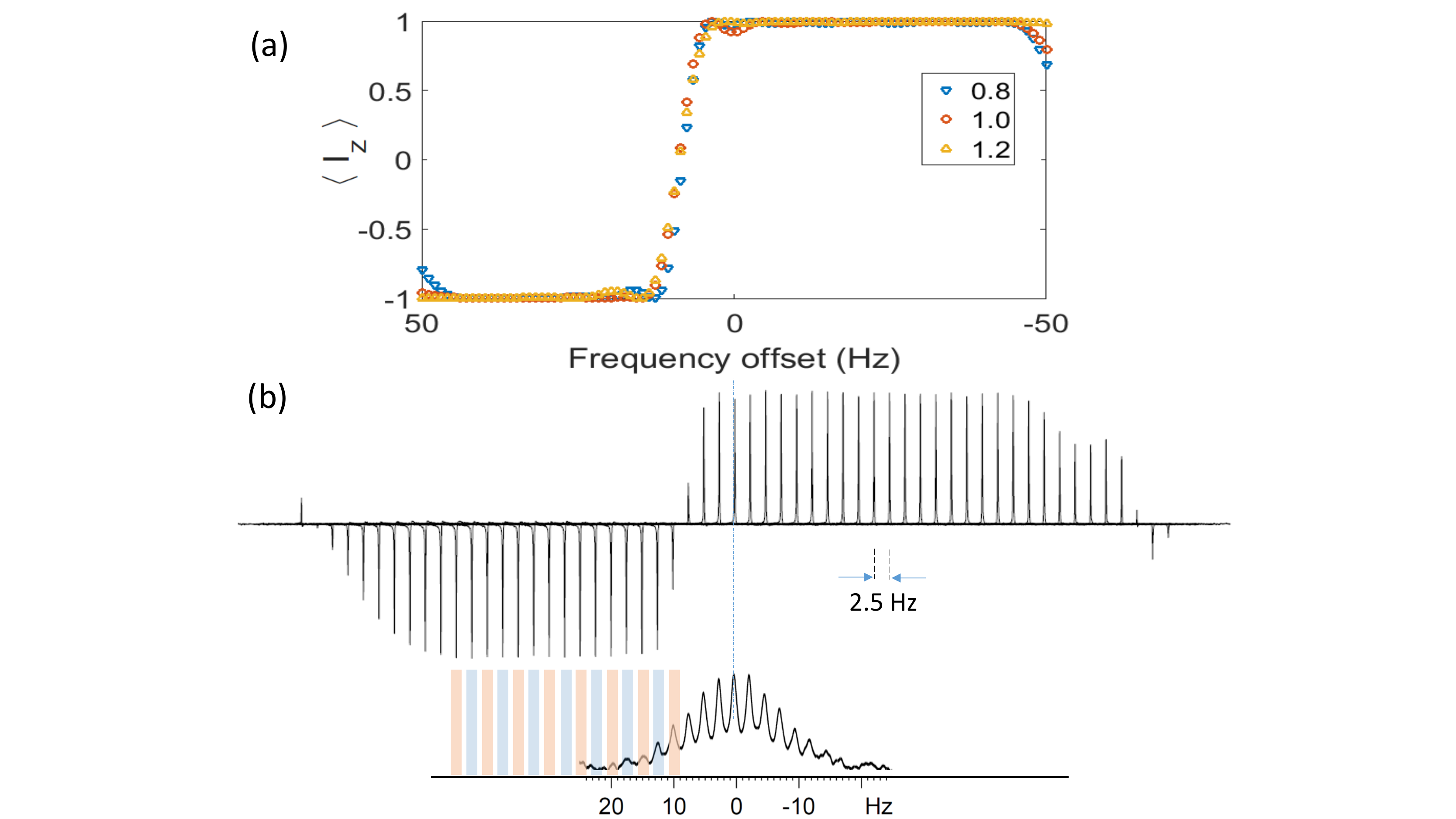}
	\caption{(a) The inversion profile of the single-step GRAPE pulse.  To illustrate its robustness w.r.t. RF inhomogeneity, three inversion profiles with 0.8, 1.0, and 1.2 times the nominal RF amplitudes are displayed. (b) Performance of the single-step GRAPE pulse experimentally observed by shifting a single-resonance peak.  The reference spectrum of TTSS is shown at the bottom. First 15 lines of TTSS are inverted since $m=15$ (see main text).}
	\label{grape}
\end{figure}

\begin{figure}[h]
	\hspace*{-0.2cm}
	\includegraphics[trim=5cm 4cm 5cm 2cm, clip=true,angle=0,width=7cm]{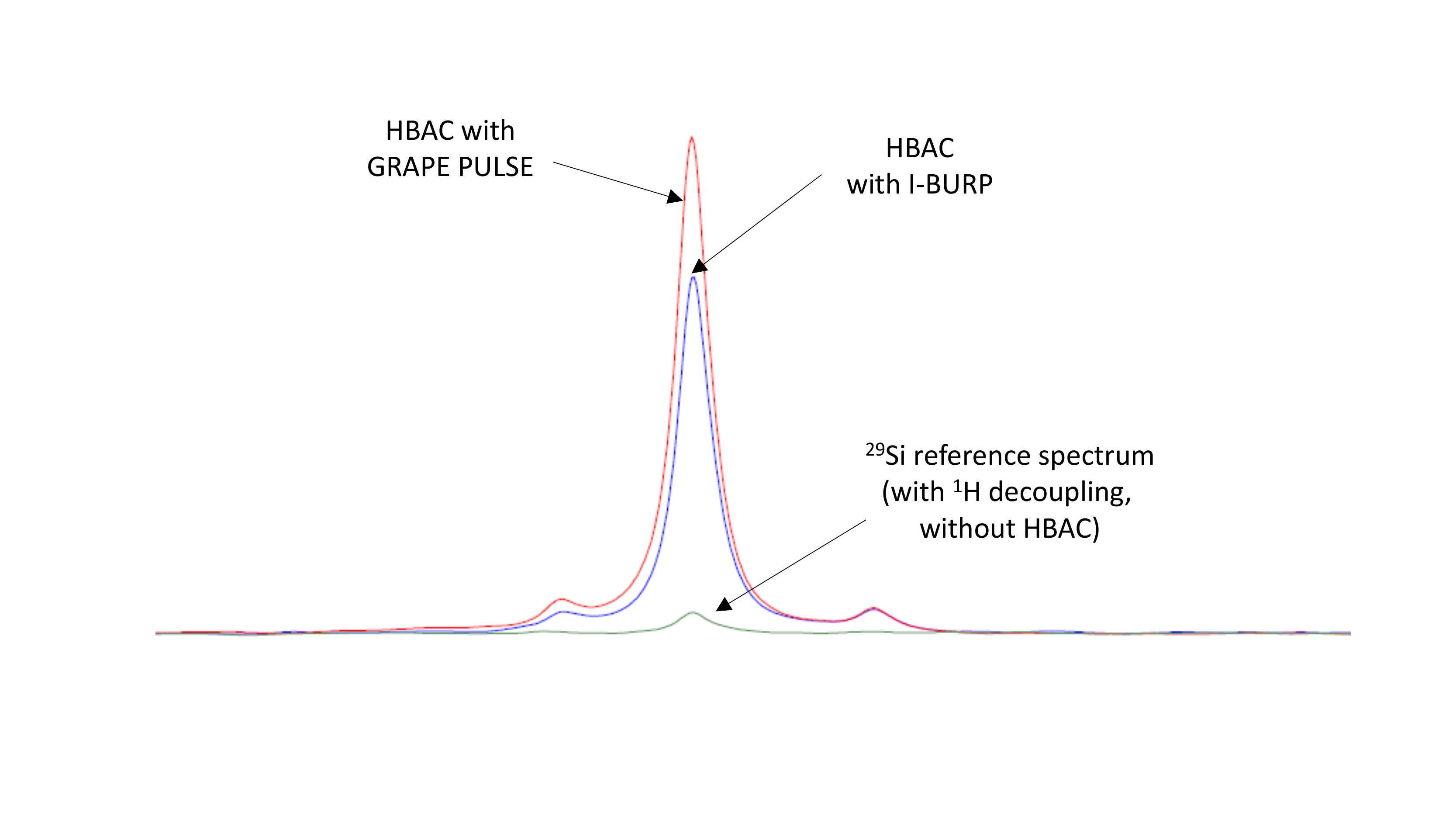}
	\caption{Comparing the HBAC performance of the single-step GRAPE pulse with IBURP-1 \cite{geen1991band}.}
	\label{grapevsiburp}
\end{figure}

\begin{figure}
	\hspace*{-0.2cm}
	\includegraphics[trim=6cm 1cm 7cm 0cm, clip=true,angle=0,width=7cm]{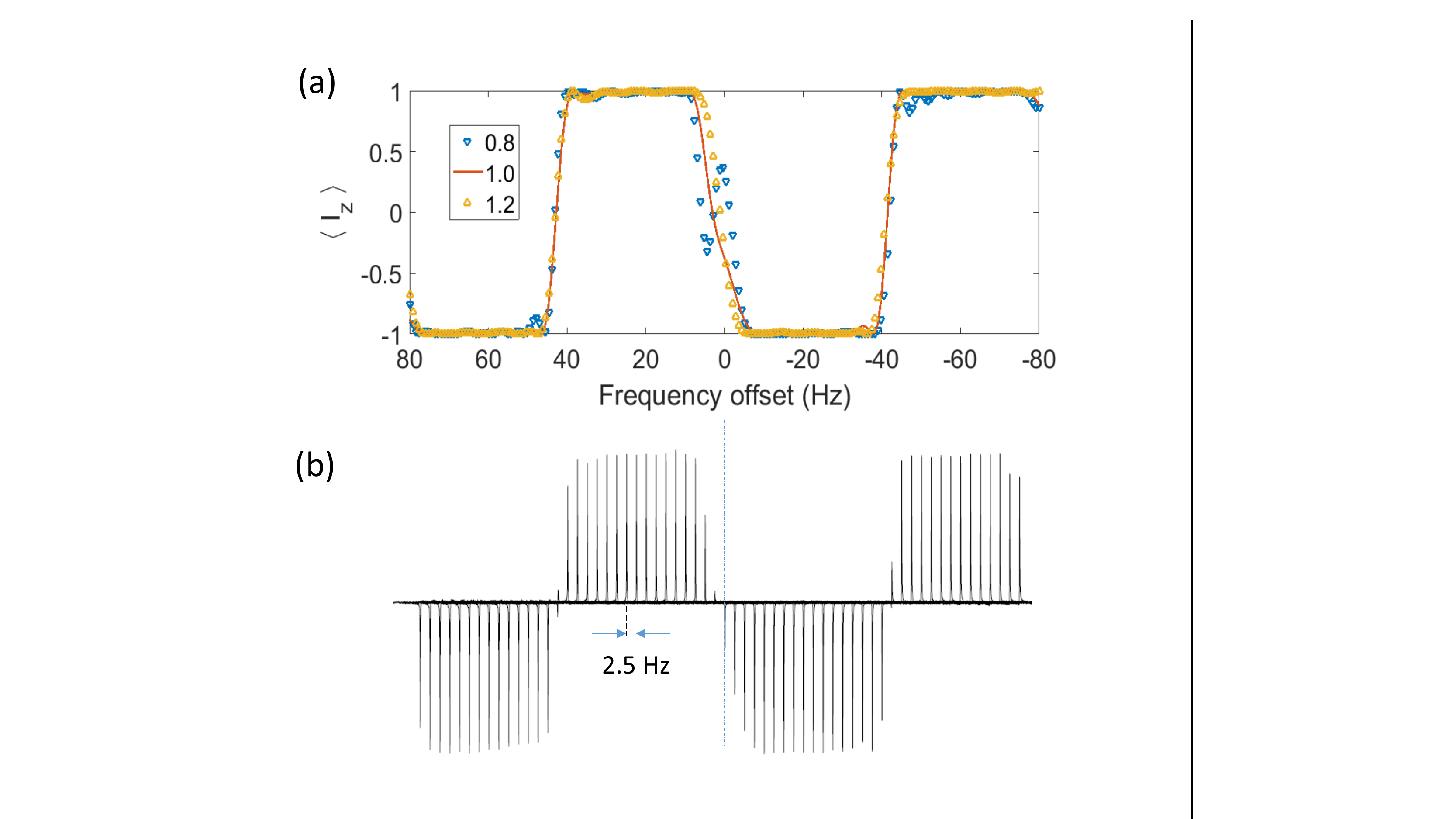}
	\caption{(a) The inversion profile of the double-step GRAPE pulse.  To illustrate its robustness w.r.t. RF inhomogeneity, three inversion profiles with 0.8, 1.0, and 1.2 times the nominal RF amplitudes are displayed. (b) Performance of the double-step GRAPE pulse experimentally observed by shifting a single-resonance peak.}
	\label{sisipulse}
\end{figure}

\section{Specially designed RF controls}
\label{appB}
For AC, we need to invert only the top-half of the populations, i.e., only the positive transitions of the computation qubit.
In order to design such a pulse, we assumed a set of about 50 two-level systems whose resonance offset are uniformly spread over a range from -50 Hz to 50 Hz.  An RF pulse for the selective inversion of half the transitions was then numerically obtained by modifying the GRAPE algorithm \cite{khaneja2005optimal}.  The  inversion profile of the single-step GRAPE pulse and its experimental performance-analysis are shown in Fig. \ref{grape}.

Although one can use standard shaped-pulses such as I-BURP to achieve band-selective inversion, we observed better performance with the robust GRAPE pulse generated by us (see Fig. \ref{grapevsiburp}).

In the case of HBAC of a pair of qubits, 
each $^{29}$Si transition is split by  $^{29}$Si-$^{29}$Si coupling followed by $^{29}$Si-$^1$H coupling.  Therefore, to cool one of the $^{29}$Si qubits (say left one as indicated in Fig. \ref{twoq} in the main text)
we need to invert two sets - each of $m$ transitions - corresponding only to top $m$ energy levels of $^1$H.  This requires a pair of inversion and no-inversion zones. Using the pulse-shaping methods described above we designed such a double-step GRAPE pulse (see Fig. \ref{sisipulse}).

\bibliography{algocool}{}

\bibliographystyle{ieeetr}

\end{document}